\documentclass{article}\usepackage[]{graphicx}\usepackage[]{color}
\usepackage{geometry}
\usepackage[parfill]{parskip}
\usepackage{graphicx}
\usepackage{amssymb}
\usepackage{epstopdf}
\usepackage{natbib}
\usepackage{color}
\definecolor{dark-gray}{gray}{0.3}
\usepackage[colorlinks=true,urlcolor=dark-gray,breaklinks,citecolor=dark-gray,linkcolor=dark-gray]{hyperref}
\usepackage{url}
\usepackage{subfig}
\usepackage{units}
\usepackage{nth}
\title{Catching crabs: a case study in local-scale English conservation}
\author{William D.\ Pearse$^1$, Helen K.\ Green$^2$, and David Aldridge$^3$}
\date{\small$^1$Department of Ecology, Evolution, and Behavior, University of Minnesota, 1987 Upper Buford Circle, Saint Paul, Minnesota, 55108, USA \url{wdpearse@umn.edu}
\\$^2$Respiratory Diseases Department, Centre for Infectious Disease Surveillance and Control, Public Health England, 61 Colindale Avenue, London, UK. \url{helen.green@phe.gov.uk}
\\$^3$Aquatic Ecology Group, Department of Zoology, University of Cambridge, Downing Street, Cambridge CB2 3EJ, UK. \url{da113@cam.ac.uk}
\\Date: \today}
\IfFileExists{upquote.sty}{\usepackage{upquote}}{}
\begin{document}
\bibliographystyle{besjournals}
\maketitle

\section{Abstract}
Wells-next-the-Sea and Cromer in Norfolk (England) both rely upon
their local crab populations, since crabbing (gillying) is a major
part of their tourist industry. Compared to a control site with no
crabbing, crabs from Wells harbour and Cromer pier were found to have
nearly six times the amount of limb damage. Crabs caught by the
general public had more injuries than crabs caught in controlled
conditions, suggesting the buckets in which the crabs were kept were
to blame. Since there is much evidence that such injuries have
negative impacts on the survival and reproductive success of the shore
crab, this is taken as evidence of non-lethal injury from humans
having a population-level effect on these animals. Questionnaire data
demonstrated a public lack of awareness and want for information,
which was then used to obtain funding to produce a leaflet campaign
informing the public of how to crab responsibly. All data collected is
available online at
\url{http://dx.doi.org/10.6084/m9.figshare.979288}.

\section{Introduction}
The Wells-next-the-Sea and Cromer economies are boosted by tourists
and locals catching crabs (`crabbers'), temporarily keeping them in
buckets and then returning them to the sea (`crabbing', also known
locally as `gillying'). While this activity requires only the purchase
of a line and bait, the crabbers (typically tourists visiting for a
day) use other facilities in the local area and so provide a steady
source of income for local businesses during the summer months. The
crabs themselves are super-abundant, simple to catch, and easy to check
for damage, while the economic effect of crabbing can be rapidly
measured with questionnaires. Thus these towns provide perfect case
studies of a problem that is seen time and time again in conservation:
a natural resource that is being damaged by the very economy that
depends on it.

The shore crab (\emph{Carcinus maenas}, Linaeus 1758) has been
intensively investigated and is considered one of the world's one
hundred most invasive species by the IUCN \citep[the World
Conservation Union, see][]{Lowe2000}. They are an intertidal species,
and while adults are occasionally stranded the majority avoid leaving
the water for extended periods of time, especially the smaller
($<35$mm) individuals that rarely leave the upper intertidal zone
\citep[][and references therein]{Naylor2003}. Males are known to
compete for females, which usually cluster in `hot-spots', although
matings outside these areas (often with smaller males) have been
documented \citep{Meeren1994}. Males frequently fight and are more
likely to do so to determine access to females
\citep{Sneddon2003}. Chela (claw) strength predicts the outcome of
fighting more accurately than size \citep{Sneddon2000}, although
relative chela size determines the likelihood of a fight being
initiated \citep{Sneddon1997b}. Fighting is intense from the very
beginning, and is more violent in the presence of food
\citep{Sneddon1997a}. There is a sexual dimorphism in chela anatomy
which cannot be explained by different feeding ecology and is likely
related to this intraspecific competition \citep{Spooner2007}. Males
frequently autotomise their limbs or sustain injury as a result of
these contests \citep{Sneddon2003}, though it should be noted that
this rarely occurs if one male is attempting to directly intervene in
the copulation of another \citep{Abello1994}.

Autotomy \citep[``\emph{a reflexive response to injury or its threat
  that results in the casting off of an appendage at a predetermined
  breakage plane}'';][]{Juanes1995} has long been documented in crab
species and is often used as an escape strategy
\citep[\emph{e.g.},][]{Wasson2002}. The limb can be regenerated in
subsequent moults as long as the crab has not undergone its final molt
(reached terminal anecdysis), and the likelihood of a crab shedding a
limb is related to its ability to regenerate in some species
\citep{Carlisle1957}. Chela loss can have consequences for mating
success \citep{Abello1994,Sekkelsten1988}, the strength of remaining
and regenerated pincers later in life \citep[causing a change in
feeding behaviour;][]{Brock1998}, moult increment \citep{Lowe2000},
and survival \citep{Smith1995}. While the loss of other locomotive
limbs can be damaging, this depends in part upon the specifics of the
loss, such as whether the loss is symmetrical \citep[discussed with
empirical data in][]{Smith1995}. Maintaining chela strength is so
important that damage avoidance has been invoked to explain an
apparent sub-optimality in the feeding ecology of the shore crab
\citep{Smallegange2003}. While some workers have drawn attention to
the difference between autotomy and the involuntary removal of a limb
\citep[\emph{e.g.},][]{Wasson2002}, arguably for our purposes this
distinction does not matter since the above studies measured simply
the effect of missing limb(s) and not autotomy \emph{per se}.

In the wider literature, the idea that non-lethal injury to organisms
can regulate their population dynamics has been modelled and
investigated to some extent in various species \citep[][and references
therein]{Harris1989}. We propose that crabbing has non-lethal impacts
on crabs in two ways: (1) by increasing fighting intensity and
frequency due to proximity in the buckets, and (2) the physiological
stress of being removed from the sea. Shore crabs have difficulty
dealing with prolonged hypoxia and osmotic stress, and tolerance
varies within a given population \citep[][and references
therein]{Reid1997}. In low oxygen conditions the outcome of aggressive
encounters are determined primarily by carapace (shell) diameter
\citep{Sneddon1998}, indicating that in these conditions the limiting
factor is aerobic capacity, not strength.

This study quantifies the effect of crabbing by measuring limb and
pincer loss frequency at a control site versus other areas of high
crabbing intensity. To determine the cause of crab injury, the crabs
inside the crabbers' buckets were compared with the crabs we caught
ourselves. While doing this we were able to assess the willingness of
the general public to change their methods of crabbing, as well as
their general knowledge of crabs to plan conservation interventions.

\section{Methods}
\subsection{Study areas}
\begin{figure}
\begin{center}
\includegraphics[width=0.5\textwidth]{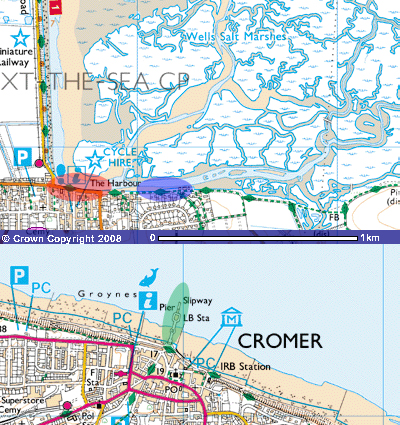}
\end{center}
\caption{Study sites. Wells-next-the-Sea at top (grid reference TF 915
  435), Cromer at bottom (UK grid reference TF 915 435). Study sites
  are shaded: red---Wells harbour, blue---control,
  green---Cromer. Images produced from the Ordnance Survey Get-a-map
  service. Images reproduced with kind permission of Ordnance Survey
  and Ordnance Survey of Northern Ireland.}
\label{map}
\end{figure}

Three study areas were investigated: Wells harbour, a nearby control
area, and Cromer pier. All three are in the county of Norfolk,
England, and are shown in detail in figure \ref{map}.

No noticeable crabbing has taken place at the control site in recent
memory (Dr David Aldridge pers.\ obs.), perhaps because it is some
distance from the main harbour and the town's amenities. The maximum
tidal range is 3.9 m (Wells harbour master pers.\ coms.) and the tidal
currents can be extremely strong (one of our secured traps was washed
away). There is little, if any, boat traffic and no boats are moored
in the area.

Wells harbour is an active fishing quayside, with numerous boats
moored alongside it throughout the day. The tide is of the same range
and strength to the control site and some areas are fully exposed at
low tide. Crabbing activity is intense during the summer months. No
dedicated facilities are available for crab catchers on the quayside,
but a freshwater tap is nearby. There has been a pier at Cromer for
over one hundred years and the area is renowned for crabbing. The pier
is purely for tourists and so there are facilities and shelter for
crabbers. It is around one hundred meters in length and at its
furthest point from the shore it is over twenty meters from the
surface of the sea. There is no significant boat traffic in the
immediate vicinity.
\subsection{Trap surveys}
Crab traps were thrown into the water with a fresh rasher of streaky
smoked bacon tied inside as bait, secured to the quayside at a marked
point with ample twine such that they came to rest on the harbour
bottom. The traps were 45 cm in length, 25 cm wide and made of 5 mm
diameter thin gauge aluminium mesh, with two holes that allowed the
crabs access to the central chamber (through a narrowing tube 7.5 cm
in diameter at the end) from which they were unable to escape. Once
the traps were hauled up the carapace width, sex, and condition of the
limbs of all the crabs inside were recorded, and then the crabs
returned to the sea. Lengths were measured with calipers, while visual
observation revealed the sex \citep[from the abdominal segments,
see][]{ Crothers1967} and limb condition.

Table \ref{schedule} shows how many traps were deployed at a
particular time at each site; note the deployments were co-ordinated
with the tidal cycle. The precise locations from which the traps were
deployed were kept constant from day three onwards, and were in all
cases evenly spread across the study site. Each trap was thrown in
sequentially with an eight minute delay as an experimenter moved from
one side of each site to the other, hence the starting side was chosen
at random. Except for initial test deployments, the traps were
deployed for 45 minutes, a period of time determined by the need to
control for the tide and the logistics of the study.

\begin{table}
\begin{tabular}{p{1.3cm} p{1.3cm} p{1.3cm} p{1.3cm} p{1.3cm} p{2cm} p{1.3cm} p{1.3cm}}\hline
& \multicolumn{4}{c}{Before low tide} & & \multicolumn{2}{c}{After low tide}\\
& \textbf{-8hrs} & \textbf{-6hrs} & \textbf{-4hrs} & \textbf{-2hrs} & \textbf{Low Tide} & \textbf{+2hrs} & \textbf{+4hrs}\\\hline
\textbf{Day 1} & & \multicolumn{5}{l}{Testing for rest of day}\\
\textbf{Day 2} & & 12 @ W & 12 @ W & & 12 @ W & 12 @ W \\
\textbf{Day 3} & & 7 @ W & 7 @ W & & 7 @ W & 7 @ W \\
\textbf{Day 4} & 7 @ C & 7 @ C & & 7 @ C & 7 @ C \\
\textbf{Day 5} & & 7 @ W & 7 @ W & & 7 @ W & & \\
\textbf{Day 6} & \multicolumn{7}{l}{Data collection at Cromer}\\
\textbf{Day 7} & 7 @ W & 7 @ W & 7 @ W & 7 @ W & & \\\hline
\end{tabular}
\caption{Outline of trap deployment schedule. Each cell square represents a two hour deployment window. In each case the number of traps deployed, followed by their location (W---Wells-next-the-Sea; c---Control site), is specified.}
\label{schedule}
\end{table}
\subsection{Bucket surveys}
At Wells harbour and Cromer we systematically moved from one side of
the study area to the other, surveying as many crabbers' buckets as
possible. The time of the survey and details of the crabs were
recorded in the same way as animals from our own traps. One such sweep
was conducted during each trap deployment window in Wells, in Cromer
once an hour, and it was usually possible to survey the vast majority
of crabbers in this time. If a group was returned to, their original
survey data was omitted and the survey started again from
scratch. Aspects of crabbing effort (number and type of equipment and
bait used and time spent crabbing) were assessed within the
questionnaire survey, which is described below.
\subsection{Mark-recapture technique}
To estimate population size, from day four onwards randomly selected
crabs collected by us and the general public were marked before
release. On the first day crabs were marked with a code identifying
the source of the crab on the carapace close to each pincer using
metallic pen; since the ink rarely completely dried on a wet crab, the
next day a bee identification tag was affixed under one pincer using
quick-drying gel. The intention was to mark five crabs from each of
the seven traps per day, and to mark five crabs in crabbers' buckets
around the area of the traps. However, while 66 crabs were marked for
recapture, at Wells only 3 were recovered, a number considered too
small for reliable population estimates to be made.
\subsection{Questionnaires}
Participants in the bucket surveys, as well as some randomly selected
people, were asked to answer questionnaires about their crabbing
activities. The questions focused on what methods were being used to
catch crabs, how the crabs were being treated and what economic impact
the crab catchers were likely to have. While thermometers were
trialled to see if they would be of use in determining the temperature
of the water in which the crabs were being kept, they were found to be
unreliable and so not used after the first day. Additional questions
were added during the study to gauge the necessity of a campaign to
educate tourists and locals about the crabs, as well as whether they
would take interest in such a campaign. A copy of the questionnaire
can be found in the appendix.
\subsection{Statistical Analysis}
All analysis was conducted using \emph{R} version $3.0.2$
\citep{R2014}. Each crab's damage was modelled as a function of the
site from which the crabs were taken, the crab's sex and carapace
diameter, the duration of deployment, and the number of crabs in its
bucket or trap, using a generalised linear model with the binomial
error family and canonical (logit) link function. Crab damage was
coded as a binary variable: crabs missing at least one limb or pincer
were classed as damged (given a value of 1), all other (undamaged)
crabs a value of 0. While the model we present contains a marginally
significant term and hence is not a minimum adequate model
\citep[see][]{Crawley2013}, deletion of the only non-significant term
did not qualitatively affect the results and so we have chosen to show
the non-significant term. All \emph{R} code used is available online
\citep[\url{http://dx.doi.org/10.6084/m9.figshare.979288};][]{Pearse2014Crabs}.
\section{Results}
Over the course of the survey 2480 crabs were surveyed, 166 of which
were from the control site, 962 collected by us at Wells harbour, 1065
surveyed in crabbers' buckets at Wells harbour and 287 from buckets at
Cromer. 12.3\% of all crabs were missing one or more limbs. Two crab
traps were lost: one due to the tidal currents, the other to
entangling and subsequent crushing underneath a boat. 112 and 13
questionnaires were conducted at Wells harbour and Cromer pier
respectively.  All of this data is available online
\citep[\url{http://dx.doi.org/10.6084/m9.figshare.979288};][]{Pearse2014Crabs}.

\begin{figure}
\centering
\includegraphics[width=0.5\textwidth]{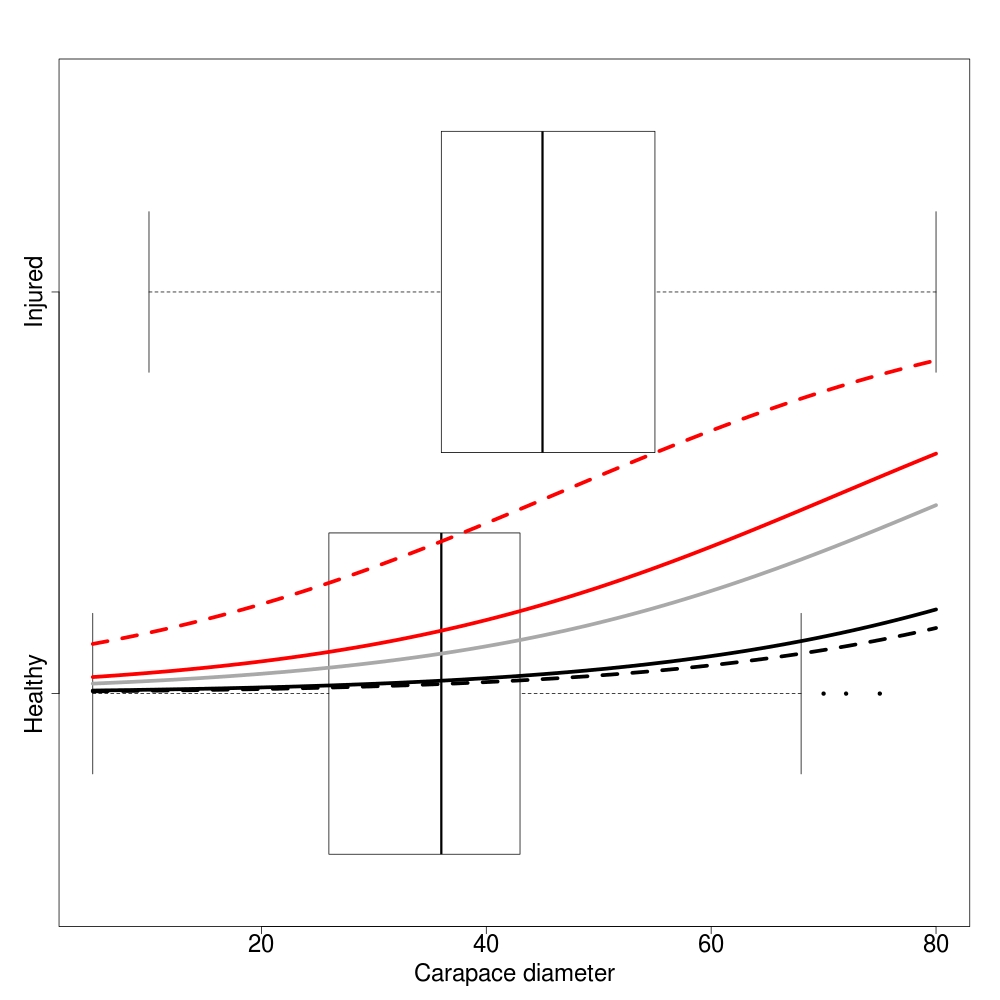} \caption[Plot of probability of injury in crabs]{Plot of probability of injury in crabs. A box-and-whisker plot of injured and healthy crabs against carapace diameter, with probability of injury added using the coefficients in table \ref{glmTable}. The black dashed and solid lines indicate probability of injury for females and males at the control site; grey indicates the probability of injury for males at Wells harbour, while the red solid and dashed lines show the probability of injury for males in tourists' buckets at Wells harbour and Cromer respectively. As shown in table 2, all lines are significantly different from one-another.\label{damagePlot}}
\end{figure}

Table \ref{glmTable} and figure \ref{damagePlot} show that crabs were
almost twice as likely to be damaged if sampled at Wells harbour in
comparison with the control site, and nearly twice as likely again at
Cromer. A crab in a tourist's bucket was almost one-and-a-half times
more likely to be damaged than if the crab were taken from our
traps. However, small crabs were relatively unlikely to be damaged in
all cases.
\begin{table}
\centering
\begin{tabular}{lrrrr}
  \hline
 & Estimate & Std. Error & z value & Pr($>$$|$z$|$) \\   \hline
Reference & -5.170 & 0.68971 & -7.496 & $>0.0001$ \\ 
Wells harbour & 1.204 & 0.60545 & 1.989 & 0.0467 \\ 
Cromer & 2.394 & 0.66381 & 3.606 & 0.0003 \\ 
Tourists & 0.518 & 0.18108 & 2.863 & 0.0042 \\ 
Time deployed & -0.003 & 0.00137 & -2.137 & 0.0326 \\ 
Male & 0.306 & 0.13705 & 2.232 & 0.0256 \\ 
Carapace diameter & 0.047 & 0.00584 & 8.093 & $>0.0001$ \\ 
Number of crabs per trap/bucket & -0.003 & 0.00113 & -2.797 & 0.0052 \\ 
-6 tide & 0.423 & 0.26187 & 1.614 & 0.1065 \\ 
-4 tide & -0.119 & 0.27198 & -0.437 & 0.6624 \\ 
-2 tide & 0.045 & 0.26411 & 0.169 & 0.8661 \\ 
+2 tide & 0.453 & 0.27577 & 1.644 & 0.1001 \\ 
+4 tide & -0.990 & 0.43181 & -2.293 & 0.0219 \\ \hline
\end{tabular}
\caption{Statistical model of crab damage, showing increased damage in tourists' buckets when location, sex, and carapace diameter are accounted for. Results from a logistic regression (binomial link; residual deviance $1624.8$ with $2409$ d.f., null deviance $1845.4$ with $2421$ d.f.), fit without the intercept to make parameter estimates easier to interpret. Note that all categorical parameters (the harbours, whether the crab is in a tourist's bucket, the sex of the crab, and the tidal cycle as described in table \ref{schedule}) are all \emph{contrasts} with respect to `reference'. `Reference' is the estimate for a female crab taken at -8 tide in the control site; thus to obtain the estimate for a male crab in a bucket at Cromer pier, the reference estimate would be added to the estimates for these factors (in this case $\approx -5.17 + 2.39 + 0.52 + 0.31 \approx -1.95$). Note that estimates are on the logit scale.}
\label{glmTable}
\end{table}

\begin{figure}
\begin{center}
\includegraphics[width=0.5\textwidth]{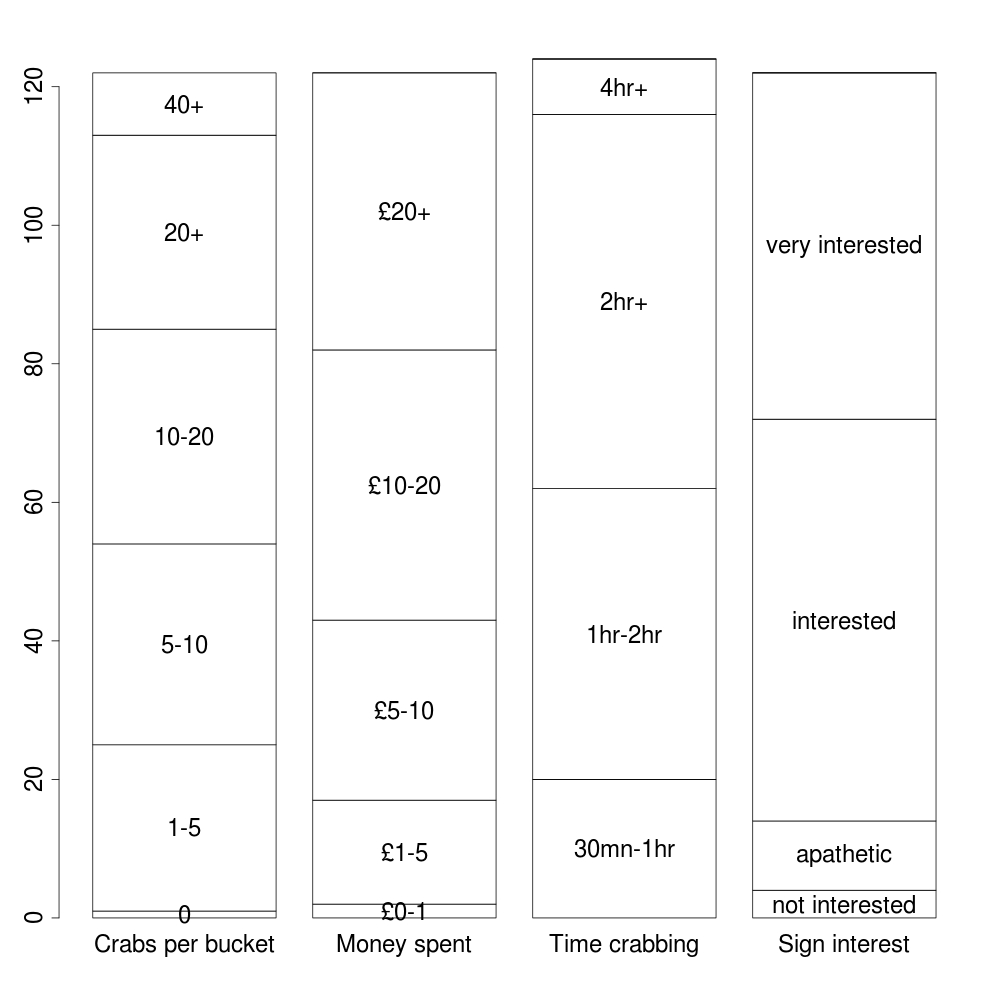}
\end{center}
\caption{Questionnaire results. Results from four survey questions;
  the vertical axis indicates the number of responses in each
  category.}
\label{survey}
\end{figure}

35\% of those asked were prepared to put more than ten crabs in a
bucket at any one time and 47\% were planning to spend longer than two
hours crabbing (figure \ref{survey}). Few people (31\% of 74 people)
knew how many legs a crab had or the species they were catching (only
25\% of 80 people answered correctly). However, 89\% would be at least
interested in the idea of a sign giving information about the crabs
(figure \ref{survey}). 12\% of people asked at Wells had freshwater
and 1\% no water at all inside their buckets, while no one asked at
Cromer used anything other than seawater.

\section{Discussion}
This study suggests that crabbing has a negative impact on crabs, as
measured through an increase in the proportion of them with missing
limbs at Wells harbour. While the study is open to criticism on the
grounds of potential pseudoreplication \citep[see][]{Hurlbert1984},
the mark and recapture study was unsuccessful precisely because we
couldn't repeatedly sample the same crabs. Indeed, given we had been
deploying the traps in the same position for several days when we
started the mark-recapture study, any behavioural bias towards the
traps would surely have been readily noticeable.
\subsection{The cause: the buckets}
The conditions in the tourists' buckets were causing stress to the
crabs in four ways: the temperature (1), salinity (2) and oxygen
content (3) of the water, and the density of crabs (4). While crabs
are naturally exposed to similar stressors as a result of the tidal
cycle \citep[described in][]{Crothers1968,McMahon1988}, they are of
much greater magnitude in the buckets and interact such that they
would likely stress the crabs more. Moreover, while crabs differ in
their exposure depending on age \citep{Naylor2003}, molt state
\citep{Reid1997} and sex \citep{Crothers1968}, these factors do not
appear to directly correlate with probability of capture.

(1) While thermometer readings were not taken, 47\% of people were
crabbing for more than two hours and there is no shade at Wells
harbour. Elevated temperature affects the acid-base balance of crabs'
blood in anaerobic conditions \citep{Truchot1973}, and although
compensation occurs within 16 hours the oxygen carrying capacity of
the blood is reduced at higher temperatures \citep{Dejours1985}. Thus
temperature could worsen an individual's reaction to hypoxia, and is
known to increase vulnerability to other stressors
\citep[\emph{e.g.},][]{Camus2004}.

(2) 12\% of the buckets in Wells contained freshwater, which is known
to be an osmotic stress for the animals. Shore crabs are not able to
regulate the osmotic pressure of their haemolymph in 20\% seawater and
certainly not freshwater \citep[][and references
therein]{Reid1997}. They show an active preference for more
concentrated seawater, distinguishing between salinity differences of
only 0.5\nicefrac{0}{00} \citep{McGaw1992}. The freshwater tap at
Wells harbour could  therefore be causing a problem.

(3) Considering the number of the crabs and the amount of time spent
in each bucket the water was likely extremely hypoxic. Even when
overcrowded, buckets of immobile crabs become noticeably more active
when fresh, oxygenated seawater is added. Shore crabs are known to
attempt to leave hypoxic water and `bubble', a mechanism by which air
is brought into the branchial cavity to supply the animal with oxygen
\citep{Reid1989}. The crabs were frequently not able to do this,
either because they were buried beneath other individuals or because
the sides of the bucket were so slippery. Tolerance to hypoxia varies
with tidal state, seemingly partly through an anticipatory mechanism
\citep{Aldrich1986}, likely leaving the crabs more vulnerable to this
stress at some times than others.

(4) 55\% of those surveyed were willing to put more than ten crabs in
a bucket at one time, conditions that often resulted in aggressive
interactions between individuals. That the smallest crabs were most
commonly found at the bottom of the buckets could be due to gravity or
fighting for access to air. In many of the more densely-packed cases
free-floating limbs could be found at the bottom of the buckets along
with a strong smell of ammonia. Males are known to be aggressive
\citep{Sneddon1997a} and the sight of a male with a female, or the
pheromone of a female, is known to make males fight more vigorously
\citep[in one study 40\% of encounters resulted in
injury;][]{Sneddon2003}. Thus close proximity should be expected to
increase the likelihood of aggressive encounters.
\subsection{The effect: missing limbs}
We therefore propose that the increase in damage in crabbers' buckets
in comparison with our crabs is the result of the conditions in the
buckets, and that this damage is the cause of the increase in injured
crabs at Wells in comparison with the control site. While there was
greater boat traffic at Wells harbour, this disturbance likely paled
in comparison to that caused by tidal currents to which the boat
traffic was related anyway. The effects of an increase in autotomy, as
outlined in the introduction, are detrimental to a crab population.

Most of the damage is likely due to fighting between crabs, which is
supported by the greater frequency of injury within a bucket
containing more crabs. That substantially more male crabs should be
injured than females is also very suggestive since they are the most
aggressive sex \citep{Sneddon1997a}. Unlike in the turbid waters of
the harbour (where the visibility was roughly 30cm), fleeing was
extremely difficult \citep[a survival strategy documented
in][]{Smith1995} and so autotomy is likely the only means of
evasion. Moreover, hypoxic conditions increase the energetic cost of
fighting \citep{Sneddon1998} such that individuals might be more
likely to autotomise for self-preservation. It is also possible that
the other stressors directly caused damage, especially if the crabs
were exerting themselves by fighting, reducing the general condition
of the crabs and so increasing the likelihood that they would need to
autotomise once released.

There are three main alternatives to this explanation. (1) The crab
catchers introduced a bias towards larger crabs (which were more
frequently injured) since they were unlikely to feel or see a smaller
crab `biting' and so wouldn't catch them. Crabbers' crabs were indeed
larger, although when size and sex are controlled for in our analysis
their crabs are still more damaged. (2) The crabs were competing
underwater for the food, such that the weaker individuals, \emph{i.e.}
those that are missing limbs, were less likely to hold onto the bait
long enough to be caught. That the public caught proportionally more
male crabs than us is highly suggestive of this, since in other crab
species males tend to win contests for resources
\citep{Briffa2007}. However, our traps would not have allowed either
the winner or loser of such a fight to subsequently leave the area,
such that if competition were taking place then the results should be
biased against the public being the source of the damage. (3) The
greater numbers of injured crabs simply reflects that each group had
been crabbing in one area longer than us and so were sampling the area
more efficiently. However, table \ref{glmTable} shows, if anything,
that traps and buckets deployed for longer had fewer injured crabs.

That Cromer should have a greater level of limb damage is difficult to
explain, but it may be due to greater levels of crabbing (Cromer
appears to have a larger tourist industry than Wells), or damage
caused to crabs as they are flung 30m from the pier into the water. We
emphasise that our sampling at Cromer was less intensive than at the
other sites, and more work would be required at Cromer.
\subsection{The effect: population-level effects}
There were no differences in sex ratio or size between the control
site and Well harbour, though there was an increase in the proportion
of males and in the size of both sexes at Cromer compared with Wells
(not shown). However, this is likely because Cromer is deeper and has
stronger currents, both factors that are known to affect the sex ratio
and average size of shore crabs \citep{Crothers1968}. Moreover, direct
comparisons between crabs caught by tourists at Cromer and Wells are
not necessarily valid since the difference in height above sea level
between the two sites could affect the catching technique. As a result
of the increased height above the sea crabbers were much less likely
to feel or see a crab on their line at Cromer unless it was much
larger, and it should be noted that male and female crabs are not, on
average, the same size.

However, it is still possible that the size and sex ratios have been
affected by crabbing and that confounding factors mask the impacts. In
aggressive encounters the strongest (and so to an extent largest) win
\citep{Sneddon1997b}, and such individuals are more likely to be red
morphs \citep{Reid1997} although these morphs are less able to cope
with hypoxic buckets \citep{Reid1989}. In contrast, smaller and weaker
animals have a lesser inter-molt period and so are able to regenerate
a lost limb more rapidly \citep{Crothers1967}. Thus limb loss may be
less costly for these individuals \citep{McVean1979}, although it
still affects smaller individuals' mating success more
\citep{Abello1994}. It is likely that a later study, perhaps looking
at inter-morph variation, would shed light on this matter.
\subsection{The solution: the importance of consultation}
Given how frequently people claimed to have been crabbing before it
was surprising that few of them seemed to even know how many legs a
crab had, let alone what species they were collecting. More
worryingly, the inability of children to recognise how many legs a
crab had suggests that they were not even looking at the crabs they
were collecting, and many seemed motivated by the size and number of
their specimens, not an interest in the crabs \emph{per se}. Thus it
may be difficult to implement measures to reduce the number of crabs
kept in each bucket and encourage them to throw the excess
back. However, it is encouraging that most of those interviewed
expressed an interest in a sign giving information about the
crabs. Perhaps the children's motivations will change when in an
environment that better facilitates learning about the animals.

\begin{figure}
\begin{center}
\includegraphics[width=0.45\textwidth]{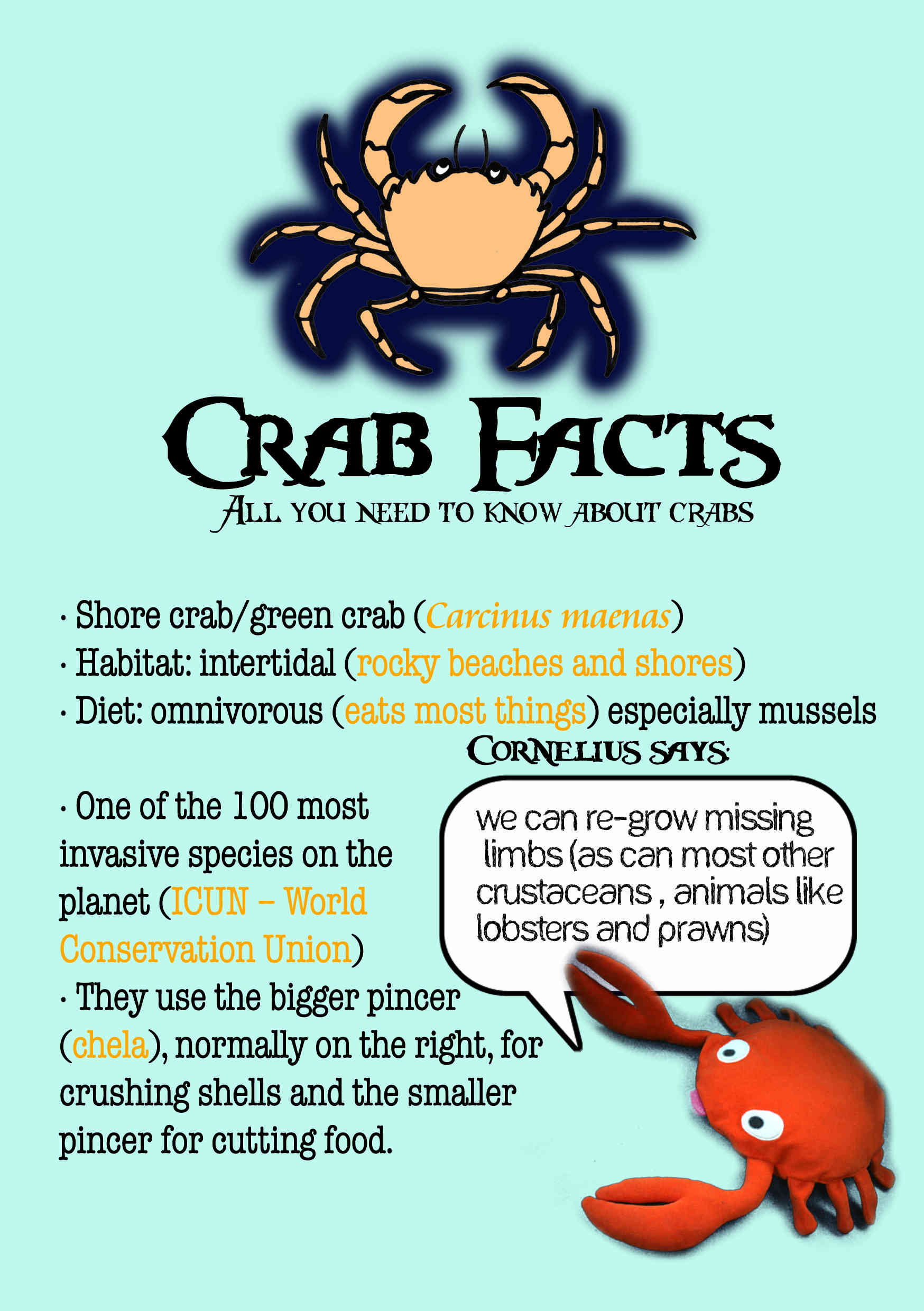}
\includegraphics[width=0.45\textwidth]{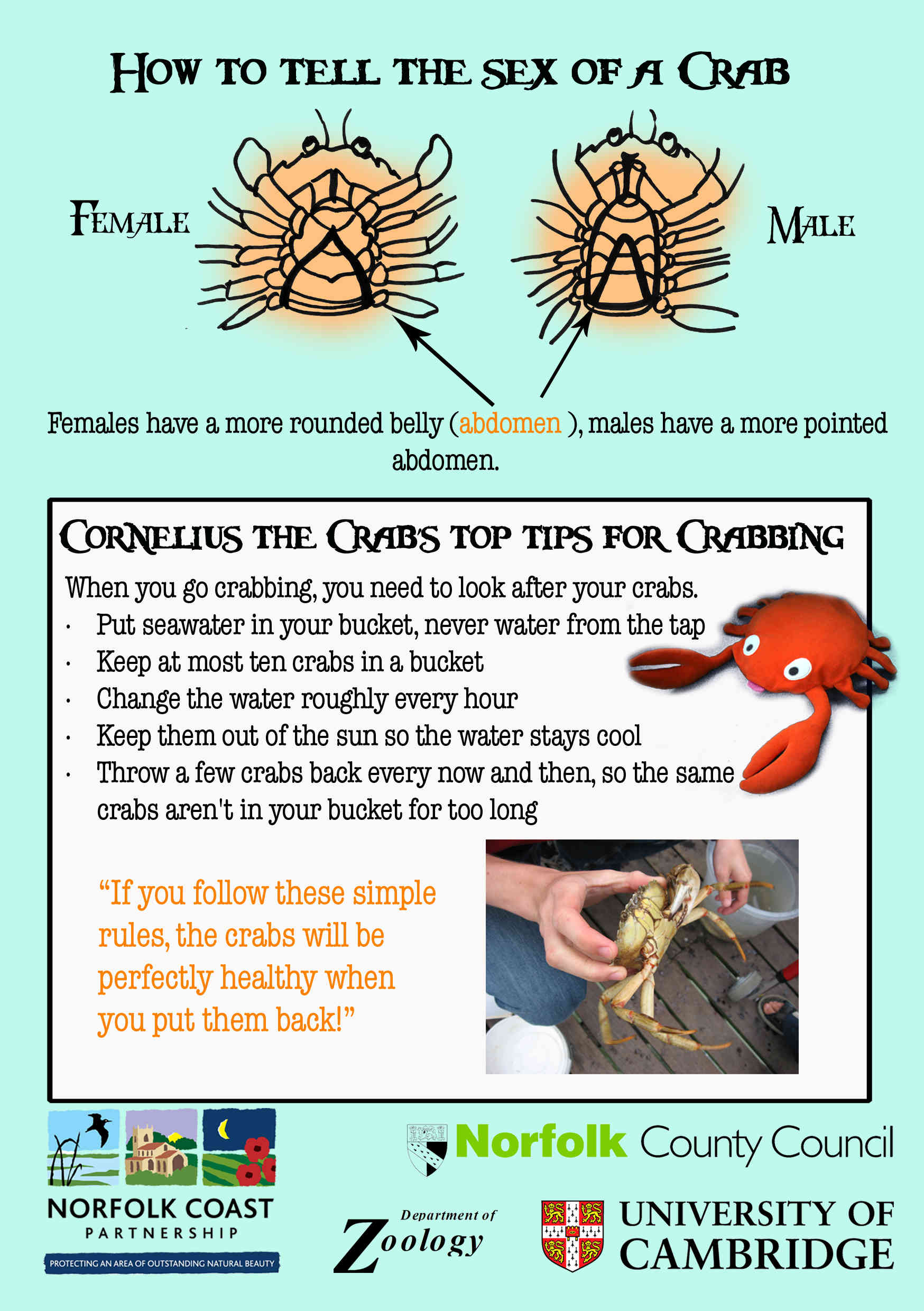}
\caption{Copy of flyer distributed to tourists as part of campaign (see text).}
\label{leaflet}
\end{center}
\end{figure}

To this end, funding was obtained from the Wells Field Center and
Norfolk Council to run a flyer campaign giving information about the
crabs and how to care for them (see figure \ref{leaflet}), in part on
the basis of the economic importance of the crabs for
Wells-next-the-Sea revealed by our questionnaire survey. The flyers
were given to bait shops in May of 2008, and a number of media outlets
covered the story (with varying degrees of sympathy for the project;
see these links:
\href{http://www.bbc.co.uk/radio2/shows/evans/moreinfo_jun08.shtml}{BBC
  Radio 2},
\href{http://www.independent.co.uk/environment/nature/the-compassionate-guide-to-catching-a-crab-843549.html}{The
  Independent},
\href{http://www.dailymail.co.uk/news/article-1025198/Council-killjoys-warn-children-dangers-crabbing--CRABS-distressed.html}{Daily
  Mail},
\href{http://www.express.co.uk/news/uk/47654/Tourist-told-take-care-when-crabbing-it-upsets-the-crabs}{The
  Telegraph},
\href{http://metro.co.uk/2008/06/09/beachgoers-told-to-be-kind-to-crabs-176857/}{The
  Metro},
\href{http://www.fakenhamtimes.co.uk/news/tips_on_looking_after_crabs_1_304588}{The
  Fakenham Times},). The public were advised to use only seawater
(reducing osmotic stress) in their buckets, change the water once
every hour (reducing hypoxia), not to leave their buckets in direct
sunlight (reducing temperature), to occasionally throw a few crabs
back and put at most ten animals in a bucket at one time (based on
preliminary analysis of these data).

This work highlights the importance of an integrated approach to any
conservation problem. Once we knew the cause of the problem, it was
simple to accumulate evidence of the gap in the public's knowledge and
of people's interest in plugging that gap. Obtaining funding was
relatively simple since we had evidence of the importance of crabbing
to the economy and had generated popular interest through the
newspaper article. The importance of generating popular and media
interest cannot be underestimated in a successful conservation
strategy. Moreover, baseline data has been accumulated (and is
available online in the supplementary materials) and the same
technique can be used to evaluate the effectiveness of the campaign,
such that it can be refined and extended if necessary. It is only
through integrating the general public into a conservation strategy
that it can be successful, for ultimately they are the ones whose
views and actions we must alter.
\section*{Acknowledgments}
This work is dedicated to the memory of Tom Pearse, who died shortly
after this fieldwork was conducted. WDP and HG played an equal part in
the data collection; this manuscript is a (very slightly) modified
version of WDP's undergraduate thesis.

\clearpage
\section*{Appendix---Questionnaire}
Due to a lack of resources, individual questionnaires could not be
printed out for each interview. Below is a copy of the questions that
each field worker asked, and after it the response types/kinds that
were recorded. As explained in the text, the questions asked changed
part-way through the study and these changes are described below.

\begin{enumerate}
\item Where do you live?
\\\emph{Local / county of home city}
\item How often have you been crabbing here in the last twelve months?
\\\emph{Integer}
\item How long have you been crabbing for so far today?
\\\emph{0-30min, 30min-1hr, 1-2hr, 2hr+}
\item How long are you planning to spend crabbing in total today?
\\\emph{0-30min, 30min-1hr, 1hr-2hr, 2hr+ (later changed to 0-30min, 30min-1hr, 1hr-2hr, 2-4hr, 4hr+)}
\item (Assessed by field worker) What equipment is being used for crabbing?
\\\emph{Line/net/hook and number of each}
\item (Assessed by field worker) What bait is being used for crabbing?
\\\emph{Bacon/sausage/sardine/any combination/etc.}
\item How much money are you planning to spend here in total today?
\\\emph{£0-£5, £5-£10, £10-£20, £20+ (later changed to £0-£5, £5-£10, £10-20, £20-£30, £30-£40, £40+)}
\item How many crabs would you be comfortable having in your bucket at any one time?
\\\emph{0, 1-5, 6-10, 10-20, 20+ (later changed to 0, 1-5, 6-10, 10-20, 20-40, 40+)}
\item What made you decide to come crabbing here today?
\\\emph{Word of mouth/tradition/impulse (i.e. passing through and saw it)/other}
\item What is inside your bucket?
\\\emph{Seawater/freshwater/nothing}
\item (Assessed by field worker) What temperature is the water in their crabbing bucket?
\item (Added later) If there were to be a sign put up, giving information about the crabs and how to crab responsibly, how interested (on the scale below) would you be in it?
\\\emph{1(not at all interested)  3(apathetic)  5 (very interested)}
\item (Added later; without the interviewee looking in the bucket) How many legs does a crab have?
\\\emph{Correct / incorrect}
\item (Added later) What is the name of the species of crab you’re catching?
\\\emph{Shore crab, green crab, \emph{Carcinus maenas} all accepted as correct responses}
\end{enumerate}

\begin{thebibliography}{34}
\providecommand{\natexlab}[1]{#1}
\providecommand{\url}[1]{\texttt{#1}}
\providecommand{\urlprefix}{URL }

\bibitem[{Abello \emph{et~al.}(1994)Abello, Warman, Reid \&
  Naylor}]{Abello1994}
Abello, P., Warman, C., Reid, D. \& Naylor, E. (1994) Chela loss in the shore
  crab \emph{Carcinus maenas} ({Crustacea: Brachyura}) and its effect on mating
  success. \emph{Marine Biology} \textbf{121}, 247--252.

\bibitem[{Aldrich(1986)}]{Aldrich1986}
Aldrich, J.C. (1986) The influences of individual variations in metabolic rate
  and tidal conditions on the response to hypoxia in \emph{Carcinus maenas}
  ({L.}). \emph{Comparative Biochemistry and Physiology Part A: Physiology}
  \textbf{83}, 53--60.

\bibitem[{Briffa \& Dallaway(2007)}]{Briffa2007}
Briffa, M. \& Dallaway, D. (2007) Inter-sexual contests in the hermit crab
  \emph{Pagurus bernhardus}: females fight harder but males win more
  encounters. \emph{Behavioral Ecology and Sociobiology} \textbf{61},
  1781--1787.

\bibitem[{Brock \& Smith(1998)}]{Brock1998}
Brock, R.E. \& Smith, L.D. (1998) Recovery of claw size and function following
  autotomy in \emph{Cancer productus} ({Decapoda: Brachyura}). \emph{Biological
  Bulletin} \textbf{194}, 53--62.

\bibitem[{Camus \emph{et~al.}(2004)Camus, Davies, Spicer \& Jones}]{Camus2004}
Camus, L., Davies, P.E., Spicer, J.I. \& Jones, M.B. (2004)
  Temperature-dependent physiological response of \emph{Carcinus maenas
  }exposed to copper. \emph{Marine Environment Research} \textbf{58}, 781--785.

\bibitem[{Carlisle(1957)}]{Carlisle1957}
Carlisle, D.B. (1957) On the hormonal inhibition of moulting in decapod
  {Crustacea II}.\ the terminal anecdysis in crabs. \emph{Journal of the Marine
  Biological Association of the United Kingdom} \textbf{36}, 291--307.

\bibitem[{Crawley(2013)}]{Crawley2013}
Crawley, M. (2013) \emph{The R Book}. John Wiley \& Sons, Ltd., \nth{2} edn.

\bibitem[{Crothers(1967)}]{Crothers1967}
Crothers, J.H. (1967) The biology of the shore crab, \emph{Carcinus maenas}
  ({L}.). 1. the background-anatomy, growth and life history. \emph{Field
  Studies} \textbf{2}, 407--434.

\bibitem[{Crothers(1968)}]{Crothers1968}
Crothers, J.H. (1968) The biology of the shore crab, \emph{Carcinus maenas}
  ({L}.). 2. the life of the adult crab. \emph{Field Studies} \textbf{2},
  579--614.

\bibitem[{Dejours \emph{et~al.}(1985)Dejours, Toulmond \&
  Truchot}]{Dejours1985}
Dejours, P., Toulmond, A. \& Truchot, J. (1985) Effects of simultaneous changes
  of water temperature and oxygenation on the acid-base balance of the shore
  crab, \emph{Carcinus maenas}. \emph{Comparative Biochemistry and Physiology
  Part A: Physiology} \textbf{81}, 259--262.

\bibitem[{Harris(1989)}]{Harris1989}
Harris, R.N. (1989) Nonlethal injury to organisms as a mechanism of population
  regulation. \emph{The American Naturalist} \textbf{134}, 835--847.

\bibitem[{Hurlbert(1984)}]{Hurlbert1984}
Hurlbert, S.H. (1984) Pseudoreplication and the design of ecological field
  experiments. \emph{Ecological Monographs} \textbf{54}, 187--211.

\bibitem[{Juanes \& Smith(1995)}]{Juanes1995}
Juanes, F. \& Smith, L. (1995) The ecological consequences of limb damage and
  loss in decapod crustaceans: a review and prospectus. \emph{Journal of
  Experimental Marine Biology and Ecology} \textbf{193}, 197--223.

\bibitem[{Lowe \emph{et~al.}(2000)Lowe, Browne, Boudjelas \&
  De~Poorter}]{Lowe2000}
Lowe, S., Browne, M., Boudjelas, S. \& De~Poorter, M. (2000) \emph{100 of the
  world’s worst invasive alien species a selection from the {Global Invasive
  Species Database}}, p.~12. {The Invasives Species Specialist Group (IUCN)}.

\bibitem[{McGaw \& Naylor(1992)}]{McGaw1992}
McGaw, I. \& Naylor, E. (1992) Salinity preference of the shore crab carcinus
  maenas in relation to coloration during intermoult and to prior acclimation.
  \emph{Journal of Experimental Marine Biology and Ecology} \textbf{155},
  145--159.

\bibitem[{McMahon(1988)}]{McMahon1988}
McMahon, B.R. (1988) Physiological responses to oxygen depletion in intertidal
  animals. \emph{American Zoologist} \textbf{28}, 39--53.

\bibitem[{McVean \& Findlay(1979)}]{McVean1979}
McVean, A. \& Findlay, I. (1979) The incidence of autotomy in an estuarine
  population of the crab carcinus maenas. \emph{Journal of the Marine
  Biological Association of the United Kingdom} \textbf{59}, 341--354.

\bibitem[{Meeren(1994)}]{Meeren1994}
Meeren, G.I.V.D. (1994) Sex- and size-dependent mating tactics in a natural
  population of shore crabs \emph{Carcinus maenas}. \emph{Journal of Animal
  Ecology} \textbf{63}, 307--314.

\bibitem[{Naylor \& Kennedy(2003)}]{Naylor2003}
Naylor, E. \& Kennedy, F. (2003) Ontogeny of behavioural adaptations in beach
  crustaceans: some temporal considerations for integrated coastal zone
  management and conservation. \emph{Estuarine, Coastal and Shelf Science}
  \textbf{58, Supplement}, 169--175.

\bibitem[{Pearse \emph{et~al.}(2014)Pearse, Green \&
  Aldridge}]{Pearse2014Crabs}
Pearse, W.D., Green, H.K. \& Aldridge, D. (2014) Catching crabs: a case study
  in local-scale {English} conservation.
  \urlprefix\url{http://dx.doi.org/10.6084/m9.figshare.979288}.

\bibitem[{{R Core Team}(2014)}]{R2014}
{R Core Team} (2014) \emph{R: A language and environment for statistical
  computing}. R Foundation for Statistical Computing, Vienna, Austria.

\bibitem[{Reid \emph{et~al.}(1997)Reid, Abelló, Kaiser \& Warman}]{Reid1997}
Reid, D., Abelló, P., Kaiser, M. \& Warman, C. (1997) Carapace colour,
  inter-moult duration and the behavioural and physiological ecology of the
  shore crab \emph{Carcinus maenas}. \emph{Estuarine, Coastal and Shelf
  Science} \textbf{44}, 203--211.

\bibitem[{Reid \& Aldrich(1989)}]{Reid1989}
Reid, D. \& Aldrich, J. (1989) Variations in response to environmental hypoxia
  of different colour forms of the shore crab, \emph{Carcinus maenas}.
  \emph{Comparative Biochemistry and Physiology Part A: Physiology}
  \textbf{92}, 535--539.

\bibitem[{Sekkelsten(1988)}]{Sekkelsten1988}
Sekkelsten, G.I. (1988) Effect of handicap on mating success in male shore
  crabs \emph{Carcinus Maenas}. \emph{Oikos} \textbf{51}, 131--134.

\bibitem[{Smallegange \& Van Der~Meer(2003)}]{Smallegange2003}
Smallegange, I.M. \& Van Der~Meer, J. (2003) Why do shore crabs not prefer the
  most profitable mussels? \emph{Journal of Animal Ecology} \textbf{72},
  599--607.

\bibitem[{Smith(1995)}]{Smith1995}
Smith, L.D. (1995) Effects of limb autotomy and tethering on juvenile blue-crab
  survival from cannibalism. \emph{Marine Ecology Progress Series}
  \textbf{116}, 65--74.

\bibitem[{Sneddon \emph{et~al.}(1998)Sneddon, Huntingford \&
  Taylor}]{Sneddon1998}
Sneddon, L.U., Huntingford, F.A. \& Taylor, A.C. (1998) Impact of an ecological
  factor on the costs of resource acquisition: fighting and metabolic
  physiology of crabs. \emph{Functional Ecology} \textbf{12}, 808--815.

\bibitem[{Sneddon \emph{et~al.}(2003)Sneddon, Huntingford, Taylor \&
  Clare}]{Sneddon2003}
Sneddon, L.U., Huntingford, F.A., Taylor, A.C. \& Clare, A. (2003) Female sex
  pheromone-mediated effects on behavior and consequences of male competition
  in the shore crab (\emph{Carcinus maenas}). \emph{Journal of Chemical
  Ecology} \textbf{29}, 55--70.

\bibitem[{Sneddon \emph{et~al.}(2000)Sneddon, Huntingford, Taylor \&
  Orr}]{Sneddon2000}
Sneddon, L.U., Huntingford, F.A., Taylor, A.C. \& Orr, J.F. (2000) Weapon
  strength and competitive success in the fights of shore crabs (\emph{Carcinus
  maenas}). \emph{Journal of Zoology} \textbf{250}, 397--403.

\bibitem[{Sneddon \emph{et~al.}(1997{\natexlab{a}})Sneddon, Huntingford \&
  Taylor}]{Sneddon1997a}
Sneddon, L.U., Huntingford, F.A. \& Taylor, A.C. (1997{\natexlab{a}}) The
  influence of resource value on the agonistic behaviour of the shore crab,
  \emph{Carcinus maenas} ({L}.). \emph{Marine and Freshwater Behaviour and
  Physiology} \textbf{30}, 225--237.

\bibitem[{Sneddon \emph{et~al.}(1997{\natexlab{b}})Sneddon, Huntingford \&
  Taylor}]{Sneddon1997b}
Sneddon, L.U., Huntingford, F.A. \& Taylor, A.C. (1997{\natexlab{b}}) weapon
  size versus body size as a predictor of winning in fights between shore
  crabs, \emph{Carcinus maenas} ({L}.). \emph{Behavioral Ecology and
  Sociobiology} \textbf{41}, 237--242.

\bibitem[{Spooner \emph{et~al.}(2007)Spooner, Coleman \& Attrill}]{Spooner2007}
Spooner, E.H., Coleman, R.A. \& Attrill, M.J. (2007) Sex differences in body
  morphology and multitrophic interactions involving the foraging behaviour of
  the crab \emph{Carcinus maenas}. \emph{Marine Ecology} \textbf{28}, 394--403.

\bibitem[{Truchot(1973)}]{Truchot1973}
Truchot, J. (1973) Temperature and acid-base regulation in the shore crab
  \emph{Carcinus maenas} ({L}.). \emph{Respiration Physiology} \textbf{17},
  11--20.

\bibitem[{Wasson \emph{et~al.}(2002)Wasson, Lyon \& Knope}]{Wasson2002}
Wasson, K., Lyon, B.E. \& Knope, M. (2002) Hair-trigger autotomy in porcelain
  crabs is a highly effective escape strategy. \emph{Behavioral Ecology}
  \textbf{13}, 481--486.

\end{thebibliography}
\end{document}